# Materials science and engineering: New vision in the era of artificial intelligence

Tao Qiang [a,*], Honghong Gao [b]

[a] School of Materials Science and Chemical Engineering, Xi'an Technological University, Xi'an 710021, China
[b] School of Mechatronic Engineering, Xi'an Technological University, Xi'an 710021, China
Correspondence to: Dr. Tao Qiang, qiangtao@xatu.edu.cn or qiangtao2005@163.com
Tel: +86-29-8617-3324, Fax: +86-29-8617-3324
Postal address: No. 2, Middle Xuefu Road, Weiyang District, Xi'an, Shaanxi, 710021, China

Materials have been central to the growth, prosperity, security, and life quality of humans since the very beginning of history. Computational materials science has gained unprecedented progress, since the concept of materials science was introduced into university teaching for common usage in USA in the 1950s. More recently, the past decade has witnessed great success and opportunities in machine-learning assisted materials discovery driven by big data and artificial intelligence (AI). Nowadays, knowledge of materials design can be created without human intervention. In this context, what possible impacts would be made to materials science and materials engineering by these emerging techniques? Furthermore, what does it mean to materials science and engineering (MSE)? To answer these questions, looking back to the history of MSE, in terms of scientific paradigm, might offer us with some new scientific insights.

Scientific discovery evolves from the 'experimental', through the 'theoretical' and 'computational', to the current 'data-intensive' paradigm, according to Jim Gray [1]. Surely, there is no exception for materials research. It means that, three paradigm shifts have occurred in the field of MSE.

Technician and craftsman had driven materials evolution principally via large-scale trial-and-error investigations for tens of thousands of years, under the first costly and time-consuming 'experimental' paradigm. Then, scientific discovery came to the second 'theoretical' paradigm with various 'laws' in the form of mathematical equations in the 1600s. In the field of materials-related science, the laws of thermodynamics are a good example [2]. This paradigm shift was proven to be a successful strategy in every aspect of physics, chemistry, and materials-related sciences.

However, some theoretical models grow too complicated to solve analytically over time. It becomes more and more difficult and costly to verify these theories with experiments. In the 1950s, modern MSE originated from the former metallurgy departments in USA. It provides the practitioners of MSE with professional education in a wide range of academic departments and programs. With the advent of computers and growth in computing power, materials scientists and engineers started computing and simulating to accelerate materials design, which advances materials innovation along experimental and theoretical approaches. Thus, materials discovery shifted to the third 'computational' paradigm in the 1960s [2]. Since then, materials scientists have developed some classical MSE models to explore the nature of MSE. The first one is materials science tetrahedron (MST) proposed in the 1980s [3]. It focuses on the basic processing-structure-properties-performance (PSPP) relationships to outline materials scenarios for its design, production, and application. The MST model advances the subject integration of metallurgy, ceramics, and polymers into an extensive MSE. In 1997, Olson proposed a central MSE model by introducing the inductive goal-means relations of engineering that combines with the deductive cause-and-effect logic of science as a systematic approach [4]. His three-link chain model, also known as 'Big Four', stresses the inverse problem for materials design. During this period, Shi declared that, composition is an equally variable distinct from structure. Processing should be linked with synthesis. Environmental effects to the property-performance relationship should be integrated into MSE models [5]. Furthermore, theory, materials and processing design, including computer simulation implicitly, should be placed at the heart of the whole enterprise. Shi presented his five-element hexahedral MSE model in 1998 [5]. To further advance the framework of MSE, Roeder and Yang took some new elements, such as length scale, function, and interface, into account their models several years ago [6,7]. Some other disciplines, such as biology, were integrated into MSE model [6]. These MSE models serve as pioneering conceptual design in hierarchically computational materials for myriad applications. As such, the past two decades has witnessed great progress in 'materials by design' [8] from first-principles calculations [9], which enables millions of materials to be high-throughput virtually screened for specific applications [10].

The contributions of computation and/or simulation to scientific progress, however, fall short of their initial promise in part, because of the extreme sensitivity of complex systems to initial conditions and chaotic behaviors [1]. MSE has accumulated mass data from computation and simulations since the 1950s, along with that from experimental and theoretical approaches. It allows scientists and engineers to design materials from a big data perspective. MSE shifted to the fourth 'data-intensive' paradigm with President Obama's launch of the Materials Genome Initiative (MGI) on June 2011.

The 'data-intensive' materials paradigm is new, beyond the experimental science, theoretical research and computer simulations. Today, MSE is on the cusp of a data revolution [11]. Due to its powerful data capture, curation, and mining potentials, big data techniques are reshaping MSE toward data-driven innovations and accelerating the materials discovery combining with AI techniques [12,13]. AI is revolutionizing materials science by decoding materials to create some potentially hypothetical candidates [14,15]. Some powerful materials data repositories, such as AFLOWLIB, Harvard Clean Energy Project, Materials Project, and Open Quantum Materials Database (OQMD) [16] have substantially boosted considerable growth in computational materials science. The MGI has provided robust solutions to data storage, sharing, and services. It has made considerable progress in materials design and property predictions. This mega project inspired some other countries, including China,

India and Japan, to launch the similar projects.

However, the trail toward data-driven MSE is clearly nontrivial, although it's prospect is really fascinating. In practice, materials scientists and engineers have been poorly equipped to deal with the deluge of large-scale complex data, while computer scientists are often ignorant of foundational theory in the materials science. The future success in data-driven MSE depends on how well we frame the right question now. It means that, materials community must articulate fresh paradigm with an innovative thinking to address the challenge put forward by big data and AI. So far, none of the previous MSE models has ever referred to these emerging techniques. Thus, a data-intensive MSE (DIMSE) model (Figure 1) was proposed to rethink, redefine, and reinvent MSE, and finally to accelerate future materials continuum.

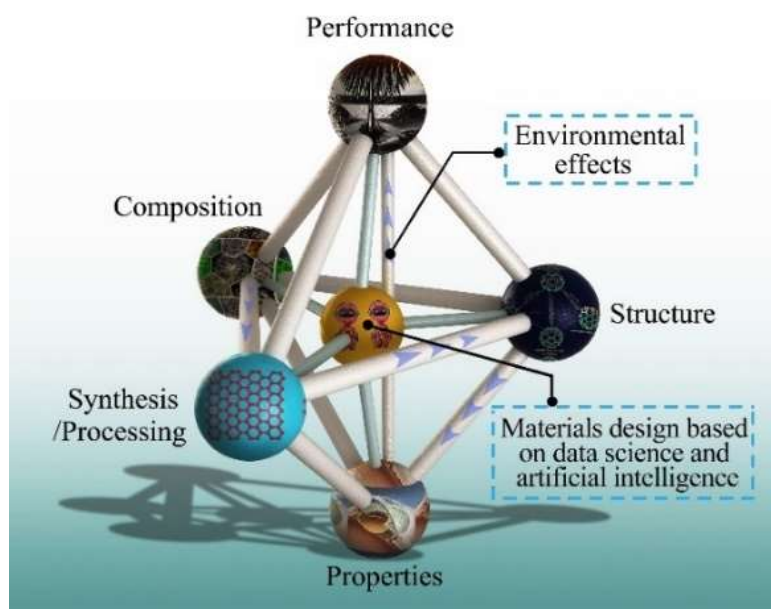

**Figure 1. The DIMSE hexahedron.** This multi-dimensional model is established to update the current MSE framework and revolutionize the next materials innovations that might become reality in the not-too-distant future.

Here, *Materials design based on data science and AI* is placed at the center of this updated hexahedron. The structure and properties of hypothetical candidate materials can be easily simulated and predicted via materials data science and AI techniques. Their compositions and synthesis/processing parameters can be designed and screened automatically. Materials scientists and engineers are encouraged to adventure the relatively simple correlations, rather than the complex causality, between the PSPP elements.

This DIMSE model provides a new scientific vision for future materials innovations. Thus, materials community can combine and advance together with big data scientists and AI experts to promote insightful cross-disciplinary exchanges of questions, methods, and results, which, in turn, will speed up future materials continuum.

4
## Acknowledgments

Thanks go to J. Chen, W.X. Chen, L. Hu, Z.Y. Jian, S.M. Wang, M.P. Wolcott, W. Yang, and H.W. Zhou for their in-depth discussion and valuable comments.

## Conflict of interests

The authors declare no conflict of interest.